\documentclass[twocolumn,numbers]{aastex631}
\usepackage{amsmath}
\usepackage{graphicx}
\usepackage{amssymb}
\usepackage{array}
\usepackage[varg]{txfonts}

\usepackage{color}
\usepackage[utf8]{inputenc}
\usepackage{comment}
\usepackage{ulem}

\makeatletter
\providecommand*{\diff}%
	{\@ifnextchar^{\DIfF}{\DIfF^{}}}
\def\DIfF^#1{%
	\mathop{\mathrm{\mathstrut d}}%
		\nolimits^{#1}\gobblespace}
\def\gobblespace{%
	\futurelet\diffarg\opspace}
\def\opspace{%
	\let\DiffSpace\!%
	\ifx\diffarg(%
		\let\DiffSpace\relax
	\else
		\ifx\diffarg[%
			\let\DiffSpace\relax
		\else
			\ifx\diffarg\{%
				\let\DiffSpace\relax
			\fi\fi\fi\DiffSpace}
\makeatother

\newcommand{\beqaa}{\begin{eqnarray}}
\newcommand{\eeqaa}{\end{eqnarray}}
\newcommand{\beqa}{\begin{equation}}
\newcommand{\eeqa}{\end{equation}}
\newcommand{\ba}{\begin{array}}
\newcommand{\ea}{\end{array}}

\def\d{\text{d}}
\def\dis{\displaystyle}
\def\({\left(}
\def\){\right)}
\def\[{\bigg[}
\def\]{\bigg]}

\begin{document}

\title{Implication of Jet Physics from MeV Line Emission of GRB 221009A}

\correspondingauthor{Zhuo Li, Zhen Zhang, Haoxiang Lin, Shao-Lin Xiong}
\email{zhuo.li@pku.edu.cn, zhangzhen@ihep.ac.cn, haoxiang@pku.edu.cn, xiongsl@ihep.ac.cn}

\author[0000-0003-4673-773X]{Zhen Zhang*}
\affiliation{Key Laboratory of Particle Astrophysics, Institute of High Energy Physics, Chinese Academy of Sciences, Beijing 100049, China}

\author[0000-0001-5729-0633]{Haoxiang Lin*}
\affiliation{Kavli Institute for Astronomy and Astrophysics, Peking University, Beijing 100871, China}

\author{Zhuo Li*}
\affiliation{Department of Astronomy, School of Physics, Peking University, Beijing 100871, China}
\affiliation{Kavli Institute for Astronomy and Astrophysics, Peking University, Beijing 100871, China}

\author[0000-0002-4771-7653]{Shao-Lin Xiong*}
\affiliation{Key Laboratory of Particle Astrophysics, Institute of High Energy Physics, Chinese Academy of Sciences, Beijing 100049, China}

\author[0000-0001-5348-7033]{Yan-Qiu Zhang}
\affiliation{Key Laboratory of Particle Astrophysics, Institute of High Energy Physics, Chinese Academy of Sciences, Beijing 100049, China}
\affiliation{University of Chinese Academy of Sciences, Chinese Academy of Sciences, Beijing 100049, China}

\author[0000-0003-1469-208X]{Qinyuan Zhang}
\affiliation{Department of Astronomy, School of Physics, Peking University, Beijing 100871, China}

\author[0000-0001-7599-0174]{Shu-Xu Yi}
\affiliation{Key Laboratory of Particle Astrophysics, Institute of High Energy Physics, Chinese Academy of Sciences, Beijing 100049, China}

\author[0000-0002-5901-9879]{Xilu Wang}
\affiliation{Key Laboratory of Particle Astrophysics, Institute of High Energy Physics, Chinese Academy of Sciences, Beijing 100049, China}

\begin{abstract} 
Ultrarelativistic jets are believed to play an important role in producing prompt emission and afterglow of gamma-ray bursts (GRBs), but the nature of the jet is poorly known owing to the lack of decisive features observed in the prompt emission. The discovery of an emission line evolving from about 37 to 6 MeV in the brightest-of-all-time GRB 221009A provides an unprecedented opportunity to probe GRB jet physics. The time evolution of the central energy of the line with power-law index $-1$ is naturally explained by the high-latitude curvature effect. Under the assumption that the line emission is generated in the prompt emission by $e^\pm$ pair production, cooling, and annihilation in the jet, we can strictly constrain jet physics with observed line emission properties. We find that the radius of the emission region is $r\gtrsim10^{16}$ cm. The narrow line width of $\sim10\%$ requires that the line emission occurs within $\sim10\%$ of the dynamical time, which further implies short timescales of pair cooling to the nonrelativistic state and pair annihilation, as well as a slightly clumpy emission region. If the jet's Lorentz factor is $\Gamma\gtrsim400$, the fast cooling requirement needs an energy density of magnetic field in the jet much larger than that of prompt gamma rays, i.e., a magnetically dominated jet. The temporal behavior of line flux suggests some angle dependence of line emission. We also discuss the difficulties of other scenarios for the observed emission line.
\end{abstract}


\section{Introduction}
It is well set up that gamma-ray bursts (GRBs) are produced by ultrarelativistic jets with bulk Lorentz factor $\Gamma\ga10^2$ in order to solve the compactness problem. However, the jet physics is largely unknown, e.g., whether the jet is kinetic- or magnetic-energy dominated and correspondingly whether the GRB radiation is powered by dissipation of kinetic energy \citep{Rees1994,Daigne1998} or magnetic field energy \citep{1997ApJ...482L..29M,2003astro.ph.12347L,2011ApJ...726...90Z}. The difficulty of the problem is that the prompt emission observed in the gamma-ray band does not provide decisive information, whereas the long-duration GRB afterglow usually loses memory on the jet physics, except for the total energy and jet opening angle \citep[see, e.g.,][for GRB review]{zhang19}.

Characteristic spectral features in GRB prompt emission are very helpful for understanding the physics of GRBs and the relativistic jet. For instance, the thermal component in GRB spectra is predicted in matter-dominated models, unless the photosphere is dissipative \citep{1986ApJ...308L..43P,2000ApJ...530..292M,2005ApJ...628..847R,2006ApJ...642..995P}. The (non)detection of thermal spectra had lead to discussion of the GRB jet composition \citep[e.g.,][]{2009ApJ...700L..65Z}.
Moreover, gamma-ray lines are expected in the GRB prompt emission, which will provide an unprecedented probe for GRB jets. During the prompt emission, a large number of electron-positron pairs are expected to form naturally in the emission region through the process of two-photon annihilation \citep{1998ApJ...494L.167P,Peer:2003yot}, 
providing a prediction for the gamma-ray emission line in GRBs from the electron-positron pair annihilation \citep[e.g.,][]{Peer:2003yot,Murase2008}.
Pair annihilation emission was also predicted at the transparency point of an opaque pair/photon plasma for GRBs \citep{Ruffini:1999ta,Ruffini:2000yu,Bianco:2001fw}.

GRB 221009A is the brightest GRB ever observed \citep[e.g.][]{Insight-HXMT:2023aof,Frederiks2023,LHAASO23,Lesage_2023,Burns_2023}, where the record-breaking isotropic energy, $E_{{\rm iso}}\sim1.5\times10^{55}$erg, is derived from the accurate measurement by GECAM-C \citep{Insight-HXMT:2023aof}. Some emission line features from about 12 to 6 MeV are reported in GRB 221009A with Fermi Gamma-ray Burst Monitor (GBM) data \citep{Maria2023} but only in the less-bright period, since Fermi/GBM suffers data issues in the bright region of this burst. Meanwhile, joint temporal and spectral analyses of GRB 221009A with GECAM-C, Fermi/GBM, and Insight-HXMT data, which were initiated soon after the detection of this burst, have led to findings of interesting spectral excess \citep{YQ2024}, as well as peculiar spectra and fine features of jet break in the early afterglow \citep{Zheng:2023xsl}. With sophisticated background modeling, careful studies of instrumental effects, and cross calibration between GECAM-C and Fermi/GBM, \cite{YQ2024} unambiguously identified these spectral excesses as line emission from the GRB by the discovery of the power-law temporal evolution of the line central energy and flux. They also found that the line energy is observed up to about 37 MeV in the early, bright part of this burst, the highest-energy emission line detected so far.

In this work, we discuss interpretations of the MeV emission line in GRB 221009A, with emphasis on the high-latitude curvature effect \citep{2000ApJ...541L..51K}.
This Letter is organized as follows. The main features of the line emission are summarized in Section \ref{sec:obs}. The physics picture of the model and assumptions are given in Section \ref{sec:model}. Then Section \ref{sec:constraint} presents analyses of observations with models and give constraints on the emission region of the MeV line. We provide discussions in Section \ref{sec:discuss}, including the difficulties in the other scenarios to account for the line emission. Finally, a summary is given in Section \ref{sec:sum}.

\section{Observation}
\label{sec:obs} 

\begin{figure}
\centerline{
\includegraphics[width=1\columnwidth]{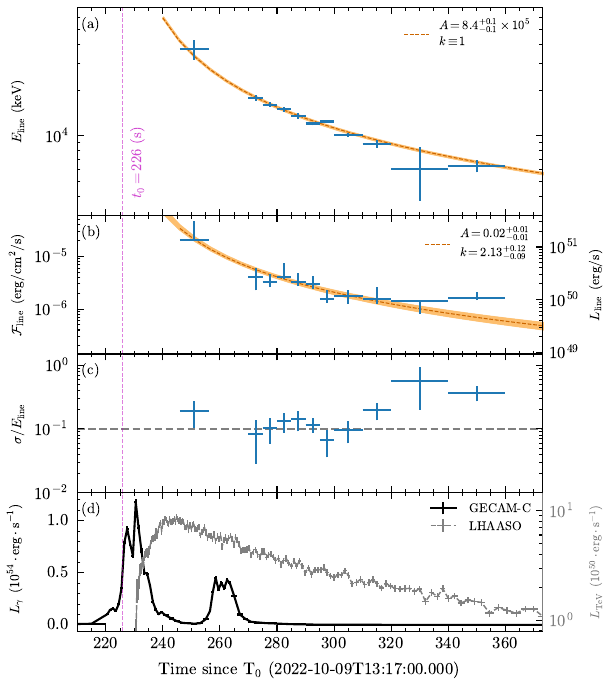}
}
\caption{The time evolution of the emission line observed in GRB 221009A, which is reproduced from the discovery paper \citep{YQ2024}. 
Panel (a): Central energy of emission line and the power law fit. Vertical purple dashed line is the best-fit initial time ($t_0$) of the power law evolution. Under the high-latitude effect framework, $k=1$ is fixed in the fitting to derive the amplitude parameter $A$.
Panel (b): Observed line flux and isotropic line luminosity with the power law fit.
Panel (c): Ratio of line width to line central energy. The horizontal dashed line denotes the 10\%.
Panel (d): Isotropic gamma-ray luminosity measured by GECAM-C (black line) and TeV luminosity observed by LHAASO (gray line).
}
\label{fig:evo}
\end{figure}

For detailed observational results of the MeV emission line from GRB 221009A please refer to \cite{YQ2024}. 
Here, we just summarize the main results (as shown in figure \ref{fig:evo}) with some dedicated analysis. 

According to the GECAM-C observation, the prompt emission of GRB 221009A mainly consist of two bright bumps at $t\sim220-240$s and $t\sim255-270$s, where $t$ is the observer time since the reference time $T_0$ which is set to 2022-10-09T13:17:00.000 UTC. The earliest detection of the emission line is at $t\sim245-255$s when the prompt emission is getting relatively low after the first bright bump of the prompt emission. 

These emission lines could be fit with the Gaussian profile, and the central energy $E_{\rm line}$ of the line evolves from $\sim37$ MeV at $t\sim245$ s to $\sim6$ MeV at $t\sim360$ s \citep{YQ2024}. Here we fit the temporal evolution of line central energy with function  
\begin{equation}
\label{eq:f(t)}
f\left(t\right)=A\,\left(t-t_0\right)^{-k},
\end{equation}
with the three parameters $A$, $k$ and $t_0$ to be determined. For the normalization ($A$) we chose an uninformative log-uniform prior.
The best fit gives $A={1.0}^{+2.0}_{-0.6} \times 10^6\,{\rm keV\,s^{k}}$,
$k=1.05_{-0.17}^{+0.22}$ and $t_0=226^{+8}_{-10}$ s. If fixing $k \equiv 1$ and $t_0 \equiv 226$ s, the best fit is $A=8.4_{-0.1}^{+0.1} \times 10^5$ ${\rm keV\,s}$. 

Like the central energy, the line flux is also observed to decay as a power law. In terms of line luminosity, it drops from $\sim 10^{51}~{\rm erg\,s^{-1}}$ to $\sim10^{50}~{\rm erg\,s^{-1}}$ during the time period of detection. The time evolution of line flux or luminosity can be fit by a power law with index $k=2.13_{-0.09}^{+0.12}$. The line width is quite narrow, and the relative line width almost keeps a constant, $\sigma/E_{\rm line}\sim10\%$, throughout the observation. 

We note that the last two data points (from 320 s to 360 s) show somewhat of a deviation from the trend of the fit. Nevertheless, the detected line emission in this time region is rather weak compared to the continuum emission and may suffer more systematic uncertainties (Figure \ref{fig:evo}). 

The duration of the time region where line emission is detected is about 115 s. However, considering the start time of the line emission $t_0\sim226$ s, the line emission is detected up to $t-t_0\sim135$ s. The line emission may be still present after 135 s but insufficient to be detectable \citep{YQ2024}.

\section{Model and Assumption}
\label{sec:model}

We assume that the emission line is produced by electron-positron pair annihilation, and the pairs are produced in the prompt phase of the GRB due to the $\gamma\gamma\rightarrow e^\pm$ process. The physical picture is such that a relativistic ejecta is released from the central engine of the GRB with a bulk Lorentz factor $\Gamma$; the prompt gamma-ray emission is produced when the ejecta expands to some distance with radius $r$, where intense pair production also occurs; if the $e^\pm$ annihilation happens rapidly, the line emission shuts down immediately, and the observed line emission should be dominated by the high-latitude effect, giving rise to decaying line energy and flux \citep{2000ApJ...541L..51K}.

\begin{figure}
\centerline{
\includegraphics[width=1\columnwidth]{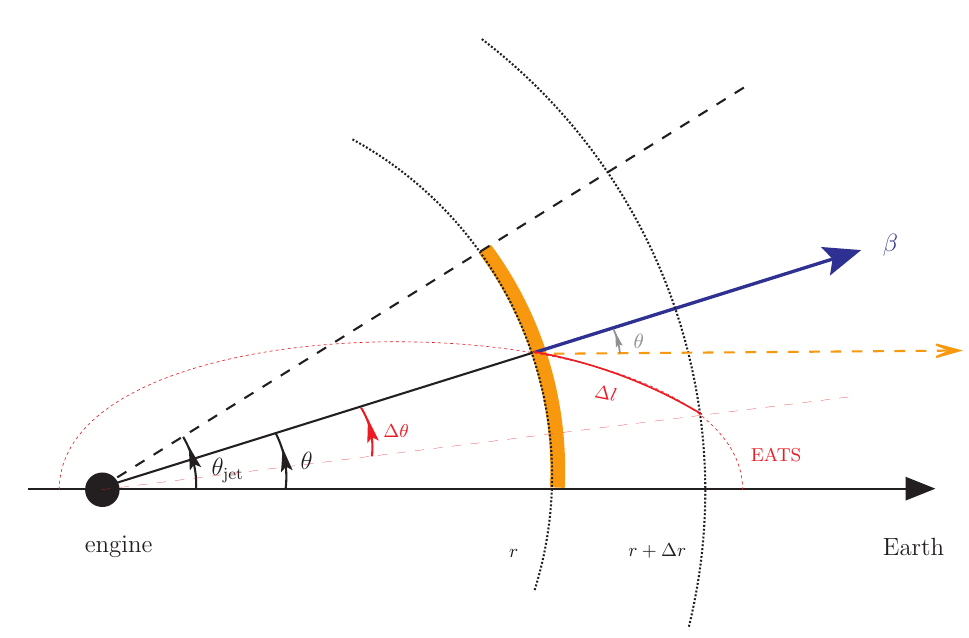}
}
\caption{Schematic plot for an expanding sphere (i.e. a jet with opening angle $\theta_{\rm jet}$) that produces emission only at the radius from $r$ to $r+\Delta r$. The EATS of a certain observer time is illustrated.
}
\label{fig:Geo}
\end{figure}

Consider an emitting spherical surface expanding outwards with Lorentz factor $\Gamma$ from the central engine at $r=0$ (Figure \ref{fig:Geo}). The ejecta may be only part of the spherical surface if the ejecta forms a jet with small opening angle.  

Let us denote that when the surface expands to radius $r$, the photon emitted by a surface element moving with an angle $\theta$ in respect to the light of sight (Figure \ref{fig:Geo}) will arrive at the observer at time $t$. Their relation is given by
\beqa
\label{eq:clocks}
t-t_0=(1+z)(1-\beta\mu)\frac{r}{\beta c},
\eeqa
where $z=0.15$ is the redshift of GRB 221009A, $\beta=(1-\Gamma^{-2})^{1/2}$, $\mu\equiv\cos\theta$, and $t_0$ is the observed arrival time of a (virtual) photon which is assumed to be emitted from the center $r=0$ at the same time as the sphere is released.  

The emitted photons are boosted by a Doppler factor $\delta_{D}\equiv1/\Gamma(1-\beta\mu)$, which, with Equation (\ref{eq:clocks}), derives
\begin{equation}
\label{eq:DFt}
\delta_{D}=\frac{r}{\Gamma}\frac{1+z}{\beta c}(t-t_0)^{-1}.
\end{equation}
Note $\beta\approx1$ for $\Gamma\gg1$. Thus, if the emission only occurs at a certain $r$ as a $\delta$-function of time in the source frame, then the emission observed at $t$ should be emitted from a certain angle $\theta=\theta(t)$ (Equation \eqref{eq:clocks}), and in observation the Doppler factor evolves as $\delta_{D}\propto(t-t_0)^{-1}$. 

If the line emission energy emitted per unit solid angle comprised by the sphere in the source frame is $\epsilon^*=dE/d\Omega$, the observed flux of the line emission is \citep[see, e.g.,][]{2000ApJ...541L..51K,Salafia2015}
\beqa
\label{eq:Liso}
{\cal F}=\frac{c}{2\,d^2_{\rm L}}\frac{\delta_{D}^3\, \epsilon^*}{r\,\Gamma },
\eeqa
where $d_{\rm L}$ is the luminosity distance of the GRB. In the simple case of constant $\epsilon^*$, the flux decays as $\propto\delta_{D}^3\propto(t-t_0)^{-3}$. If $\epsilon^*$ is $\theta$-dependent, the decay index may deviate from $k=3$.

\section{Constraints of Model by Observation}
\label{sec:constraint}

We make constraints on the physical condition of the jet in which the model can explain the observational features of the line emission, in particular the values of $r$ and $\Gamma$ of the emission region where the pair annihilation line is generated. 

\subsection{Line Energy and Evolution}

Note, the best fit of the line energy evolution gives $k\approx1$, well consistent with the prediction of the high-latitude effect (Equation\,\ref{eq:DFt}). This strongly suggests the high-latitude effect at work, whereas other scenarios usually cannot avoid fine-tuning (see the discussion in Section \ref{sec:discuss}). Moreover, the best-fit value $t_0=226$ s supports that the ejecta producing the line emission is the same as that accounting for the first bump of the prompt emission.

Assuming the line is generated by pair annihilation where the pairs are nonrelativistic (NR) in the comoving frame of the sphere (see the discussion in Section \ref{sec:width}), the line energy in the comoving frame is $\sim m_ec^2=0.511$ MeV. The Doppler factor can be given by 
\begin{equation}
    \delta_{D}= (1+z)E_{\rm line}/m_{e} c^{2},
\end{equation}
with $E_{\rm line}$ observed to be $\sim37$ MeV in early time, corresponding to $\delta_D\sim80$. This sets a lower limit of the jet Lorentz factor $\Gamma>\delta_D$, if $\theta>1/\Gamma$.

Comparing Equation \eqref{eq:DFt} with observation, one can strictly constrain $r/\Gamma$ from the fit of line energy evolution. Given the best-fit value $\log A=5.92$ and $\beta\approx1$, one has
\beqa
\label{eq:r.vs.Gamma}
r= 2.45\times10^{16}~{\rm cm} ~\(\frac{\Gamma}{500}\).
\eeqa
We also show this constraint from line energy evolution in Figure \ref{fig:r-Gamma}. It is interesting to note that the dynamical timescale in the comoving frame\footnote{Throughout this Letter, the quantities in the comoving frame are denoted with a prime sign.} 
could be firmly determined by the observed evolution of line central energy: $t^{\prime}_{\rm dyn}\simeq r/\Gamma c=A/m_{e} c^{2}=1.64\times10^{3}$ s.

The line emission should firstly come from the part of the sphere at $\theta=0$ at the observer time
\beqaa
\label{eq:tdyn2}
t-t_0=\frac{r(1+z)}{2\Gamma^2c}=1.4~{\rm s}~\(\frac{\Gamma}{500}\)^{-1},
\eeqaa
which should be the starting time of the observed power-law decay of line energy with temporal index $k=1$. 

The later line emission comes from the part with larger $\theta$, thus the time that the line emission lasts for can be used to constrain the jet opening angle $\theta_{\rm jet}$. Equation (\ref{eq:clocks}) gives $\theta=\sqrt{2c(t-t_0)/r(1+z)}$, and using Equation (\ref{eq:r.vs.Gamma}) and the lower limit of line emission duration $t-t_0\ga135$s, one has   
\beqa
\label{eq:openangle2}
\ba{rcl}
\dis
\theta_{\rm jet}\ga 0.017~{\rm rad}~ \(\frac{\Gamma}{500}\)^{-1/2},
\ea
\eeqa
which is consistent with the estimates from the jet break in early afterglow \citep{Zheng:2023xsl,LHAASO23,Insight-HXMT:2023aof}.

\subsection{Line Luminosity and Evolution} 

The initial line luminosity $L_{\rm line}$, corresponding to the part of the sphere at $\theta=0$, is not measured in the observation. The line emission should start to decrease at the time defined in Equation\,\eqref{eq:tdyn2}. The initial line luminosity should be greater than that of the earliest line detection $\sim10^{51}$erg\,s$^{-1}$ (Figure \ref{fig:evo}) and lower than the luminosity of the prompt emission $L_\gamma\sim10^{54}$erg\,s$^{-1}$. Therefore we conservatively estimate the initial line luminosity  
$ L_{\rm line}\approx 10^{52.5\pm1.5}~{\rm erg\,s^{-1}}$ and derive the constraints accordingly. 
 
In the prompt emission region, pair production is expected to be contributed significantly by photons above a cutoff energy $\epsilon_c'$ in the comoving frame ($\epsilon'\equiv h\nu'/m_e c^2$), where the optical depth is $\tau_{\gamma\gamma}(\epsilon_c') \simeq 1$. For a power-law photon number spectrum $\d N/\d\epsilon' \propto \epsilon'^{-(\alpha+1)}$, the optical depth is estimated to be $\tau_{\gamma\gamma}(\epsilon') \approx (\alpha\eta_{\alpha}/2) \sigma_T \delta N_{\gamma}|_{>1/\epsilon'}/4\pi r^2$.
Here $\eta_{\alpha}$ is a numerical factor depending on the spectral index $\alpha$ \citep{1987MNRAS.227..403S, 2001ApJ...555..540L} and typically $\eta_1=11/90$, 
the factor $1/2$ is a self-interaction correction, and $\delta N_{\gamma}|_{>1/\epsilon'} \approx   [1/(\epsilon'\epsilon_{p}')]^{-\alpha}\delta N_\gamma$ is the number of photons emitted in a single light-curve pulse above energy $1/\epsilon'$, estimated by $\delta N_\gamma$ the photon number above spectral peak $\epsilon_p'$.
In the observer frame, $\delta N_\gamma \approx \delta E_\gamma/\epsilon_p m_e c^2$, where $\delta E_\gamma$ is the observed energy released, and $\epsilon_p \approx 1$. Because of the steep GRB spectrum, only cutoff at the high-energy end is expected, $\epsilon_c'\ga1$.
Then we have 
$
    \epsilon_c = \max( \Gamma, \epsilon|_{\tau_{\gamma\gamma}=1} )
$ \citep[e.g.,][]{2010ApJ...709..525L},
where 
\beqaa\label{eq:tau=1}
    && \epsilon|_{\tau_{\gamma\gamma}=1} = \frac{\Gamma^2}{\epsilon_p}\left[\frac{4\pi r^2 \epsilon_p m_e c^2}{(\alpha \eta_{\alpha}/2)\sigma_T \delta E_\gamma}\right]^{1/\alpha}
    \\\nonumber &&= 6300 \,  
    \left(\frac{r}{10^{16}\,{\rm cm}}\right)^{2/\alpha} \left(\frac{\Gamma}{500}\right)^{2} \left(\frac{\delta E_\gamma}{10^{54}\,{\rm erg}}\right)^{-1/\alpha} \ .
\eeqaa
Here and after all numerical values are evaluated at $\alpha=1$. 
When $\Gamma=\epsilon|_{\tau_{\gamma\gamma}=1}$, we have
\begin{equation}
    r_\Gamma=0.28\times10^{16}{\rm cm}\,\left(\frac{\Gamma}{500}\right)^{-\alpha/2} \left(\frac{\delta E_\gamma}{10^{54}\,{\rm erg}}\right)^{1/2},
\end{equation}
i.e., at $r>r_\Gamma$, $\epsilon_c$ is given by Equation\,\eqref{eq:tau=1}.

The photon number outflow rate above $\epsilon_c$ is $\dot{N}_{\gamma}|_{>\epsilon_c} = (\epsilon_c/\epsilon_p)^{-\alpha}\dot{N}_\gamma$ where $\dot{N}_\gamma \approx L_\gamma/(\epsilon_p m_e c^2)$. If the pairs formed by photons above $\epsilon_c$ cool and annihilate rapidly with respect to the dynamical timescale (see the required conditions below), we estimate the emission line luminosity to be $L_{\rm line} \approx (\Gamma m_e c^2)\,\dot{N}_{\gamma}|_{>\epsilon_c} \approx (\Gamma/\epsilon_p) \, (\epsilon_c/\epsilon_p)^{-\alpha} L_\gamma$. For $\epsilon_c>\Gamma$, 
\beqaa
\label{eq:LPpm}
&&L_{\rm line} = 7.9\times10^{52}{\rm erg\,s}^{-1}\left(\frac{L_\gamma}{10^{54}{\rm erg\,s}^{-1}}\right) 
\\\nonumber && \left(\frac{r}{10^{16}\,{\rm cm}}\right)^{-2} \left(\frac{\Gamma}{500}\right)^{-(2\alpha-1)} \left(\frac{\delta E_\gamma}{10^{54}\,{\rm erg}}\right);
\eeqaa
and for $\epsilon_c\approx\Gamma$, $L_{\rm line}\sim L_\gamma$.
Given the estimated initial line luminosity $L_{\rm line}\approx10^{52.5\pm1.5}$erg\,s$^{-1}$, the constraint for $\epsilon_c>\Gamma$ is
\beqaa
    \label{eq:L_line}
    && r=2.8\times10^{16}\,{\rm cm} \, \left(\frac{\Gamma}{500}\right)^{-(2\alpha-1)/2}\\\nonumber
    &&\left(\frac{L_{\rm line}}{10^{52}\,{\rm erg\,s}^{-1}}\right)^{-1/2}\left(\frac{L_\gamma}{10^{54}\,{\rm erg\,s}^{-1}}\right)^{1/2}\left(\frac{\delta E_\gamma}{10^{54}\,{\rm erg}}\right)^{1/2}.
\eeqaa

The observed temporal index of flux deviates from $k=3$, implying a possible angle dependence of parameters of the jet or line emission. If one of the options assums $\epsilon^*\propto \theta^{2a}$, then the flux varies as $\propto\delta_D^3\,\epsilon^*\propto(t-t_0)^{-3+a}$. The observed line flux decay index of $k\approx2$ requires $a\approx1$.

\subsection{Line Width}
\label{sec:width}
The observed narrow line width of only $\sim10\%$ implies weak Doppler broadening of the line, which makes constraints on the effects of both the thermal and bulk motion.

\subsubsection{Thermal Motion}
The narrow line width requires that the pairs when annihilating with each other should be NR, with velocity $\beta'_e\sim0.1$ in the comoving frame of the ejecta.

The newly born pairs from the prompt gamma rays are energetic. They may cool down and become NR via synchrotron radiation or inverse Compton (IC) scattering off the prompt MeV photons.
In the comoving frame, the energy density of target photons for IC scatterings is estimated as
$U_{\gamma}'=L_{\gamma}/4\pi r^{2} \Gamma^2 c$. Similarly, the magnetic field energy density in the comoving frame is $U_B'=L_B/4\pi r^{2} \Gamma^2 c$, provided that $L_B$ is the magnetic field energy luminosity of the ejecta.

If denoting $\gamma_e'$ and $\beta_e'$ as the Lorentz factor and velocity of pairs in the comoving frame, respectively, the energy-loss rate is then 
\beqaa
\label{eq:gammaSR/IC}
\frac{\d \gamma_{e}'}{\d t^{\prime}} =-\frac{4}{3}\frac{\gamma_{e}'^{2}\beta_{e}'^{2}\sigma_{\rm T} cU'}{m_{e} c^2}=-\frac{\gamma_{e}'^{2}\beta_{e}'^{2}}{2 \tau},
\eeqaa
where $\sigma_{\rm T}$ is Thomson-scattering cross section, $t^{\prime}$ is the time in the comoving frame, $U'=U_B'+U_\gamma'$, and  
$\tau=3\pi r^2 \Gamma^2 m_{e} c^{2}/2\sigma_{\rm T}L_{\rm em}$, with $L_{\rm em}=L_B+L_\gamma$.
By solving Equation \eqref{eq:gammaSR/IC}, one has
\beqa
\label{eq:beta2}
\beta_{e}'^{2}=4\, e^{-t^{\prime}/\tau}\(2+e^{-t^{\prime}/\tau}\)^{-2},
\eeqa
insensitive to the initial energy of pairs. For $\beta_{e}'=0.1$, $t^{\prime}=t^\prime_{{\rm NR}}=4.6\,\tau$.
Then, for pairs to efficiently cool down to NR within a dynamical time, it is required that $t^\prime_{{\rm NR}}$ is smaller than the dynamical time, $t^\prime_{{\rm NR}}\la t^{\prime}_{\rm dyn}$. However, the observed narrow line width even requires $t^\prime_{{\rm NR}}\la 0.1t^{\prime}_{\rm dyn}$ (see Section \ref{sec:bulk-motion}), which reads $r\la\sigma_{\rm T}L_{\rm em}/69\pi\Gamma^{3}m_{e}c^3$, i.e.,
\beqaa
\label{eq:rcSR1/IC1}
r\la10^{16}{\rm cm}\(\frac{\Gamma}{500}\)^{-3}\(\frac{L_{\rm em}}{10^{55}{\rm erg\,s^{-1}}}\).
\eeqaa
Note that although in observation the peak luminosity is only $L_\gamma\sim10^{54}$erg\,s$^{-1}$, we plug in a larger electromagnetic luminosity, $L_{\rm em}=10^{55}~{\rm erg\,s^{-1}}$, here. 

\subsubsection{Bulk Motion}
\label{sec:bulk-motion}

The emission line can also be broadened by the bulk motion of the sphere if the line emission lasts for a finite duration. Consider that the line emission occurs when the sphere expands from $r$ to $r+\Delta r$, assuming $\Delta r/r\ll1$. In this case we should consider that the emission observed in an observer time $t$ is integration over a so-called equal arrival time surface \citep[EATS; e.g.,][]{1998ApJ...494L..49S}. In Equation (\ref{eq:clocks}), once fixing $t$, the $r-\theta$ relation determines an EATS for time $t$. For different angle $\theta$ the Doppler factor is different, leading to a broadening of the emission line. The part of EATS within the range $\Delta r$ has a solid angle of $\Delta\Omega={\rm sin}\,\theta\,\Delta\theta$. Performing variations in Equation (\ref{eq:clocks}) for constant $t$, we have $\Delta(1-\beta\cos\theta)r+(1-\beta\cos\theta)\Delta r=0$. Therefore, with $\delta_D$ definition, we have
\begin{equation}
\label{eq:variation1}
\frac{\Delta r}{r}=\frac{\Delta \delta_{D}}{\delta_{D}}\lesssim0.1,
\end{equation}
where the last inequality comes from the observation of narrow line width. This is consistent with the assumption $\Delta r/r\ll1$ before performing variations.

This constraint on the emission region, $\Delta r/r\la0.1$, requires that all the time scales of processes related to line emission should be $\la0.1t_{\rm dyn}'$, such as the time that pairs cool to NR, $t_{\rm NR}'$, and the pair annihilation time, $t_{\rm ann}'$. We consider the latter here.

We consider that pairs cool to an NR state before annihilation, as the cross section of pair annihilation into gamma rays increases quickly with pair velocity decreasing. When pairs become NR, $\beta'_e\sim0.1$, the cross section of annihilation is approximately 
$\sigma_{e^{+}e^{-}}\simeq\frac{3}{8}\sigma_{\rm T}\,\beta_e^{\prime-1}$.
The timescale for pair annihilation with each other is $t^{\prime}_{\rm ann}\simeq1/\sigma_{e^{+}e^{-}}n^{\prime}_{\pm}\,\beta_e^{\prime}\,c=(8/3)/\sigma_{\rm T}\,n^{\prime}_{\pm}c$, where $n_\pm'$ is the pair number density in the annihilation region in the comoving frame. 
If the spatial distribution of NR pairs is clumpy with volume filling factor $f_{\rm v}<1$,  then $n_\pm'=f_{\rm v}^{-1}\langle n_\pm'\rangle$, where $\langle n_\pm'\rangle$ is the spatially averaged number density of pairs. The pair number density can be obtained by considering the balance between formation rate density (spatially averaged) $ \dot{N}_{\gamma}|_{>\epsilon_c}/4\pi r^2 c \Gamma t'_{\rm dyn} \simeq (\epsilon_c/\epsilon_p)^{-\alpha} L_\gamma/4\pi r^3 m_e c^2 \epsilon_p$ and annihilation rate density (spatially averaged)
$\langle n_\pm'\rangle/t'_{\rm ann}\simeq (3/8) \langle n_\pm'\rangle n_\pm' \sigma_T c$.
This results in
$n_\pm' = f_{\rm v}^{-1/2} [(8/3)(\epsilon_c/\epsilon_p)^{-\alpha} L_\gamma/4\pi r^3 m_e c^3 \sigma_T \epsilon_p]^{1/2}$.
The constraints under the requirement $t^{\prime}_{\rm ann}\lesssim 0.1t^{\prime}_{\rm dyn}$ give
\beqaa 
\label{eq:ann}
 r&& \la 10^{15}\,{\rm cm}\times\min\left[ 0.065\,f_{\rm v}^{-1}\left(\frac{\Gamma}{500}\right)^{-(\alpha+2)} \!\left(\frac{L_\gamma}{10^{54}\,{\rm erg\,s^{-1}}}\right)^{}\!, \right.~~~~~~\\
 \nonumber && \left. 0.80\,f_{\rm v}^{-1/3}\left(\frac{\Gamma}{500}\right)^{-(2\alpha+2)/3} \left(\frac{\delta E_\gamma}{10^{54}\,{\rm erg}}\right)^{1/3} \left(\frac{L_\gamma}{10^{54}\,{\rm erg\,s^{-1}}}\right)^{1/3} \right].
\eeqaa

\begin{figure}
\centerline{
\includegraphics[width=1\columnwidth]{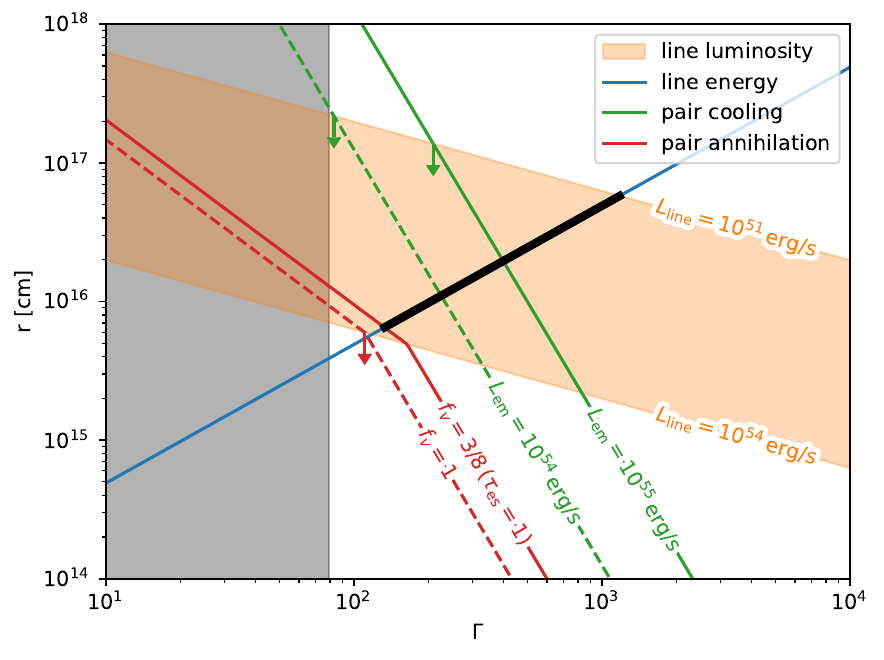}
}
\caption{Constraints of $r-\Gamma$ space for the line emission region by Equations. (\ref{eq:r.vs.Gamma}) (blue line), (\ref{eq:L_line}) (brown shaded area), (\ref{eq:rcSR1/IC1}) (green lines), and (\ref{eq:ann}) (red lines). It is assumed that $L_{\gamma}=10^{54}$erg\,s$^{-1}$, $L_{\rm em}=10^{54}$erg\,s$^{-1}$, $\delta E_{\gamma}=10^{54}$erg, and $f_{\rm v}=1$, except that $L_{\rm em}=10^{55}$erg\,s$^{-1}$ for solid green line, and $f_{\rm v}=3/8$ for solid red line.  
The gray shaded area is not allowed since $E_{\rm line}>37$ MeV gives $\Gamma>80$. The dark thick segment is the allowed parameter space for $\tau_{\rm es}<1$.}
\label{fig:r-Gamma}
\end{figure}

\section{Discussion}
\label{sec:discuss}
\subsection{Physical Implication of Constraints}

The constraints are summarized in Figure \ref{fig:r-Gamma}. The blue line and the brown shaded area together conclude that the radius of the emission region should be in the range of $10^{16}$ cm $\la r\la10^{17}$ cm, whereas the jet bulk Lorentz factor is constrained to be $10^2\la\Gamma\la10^3$. 

Comparing the blue and green lines, the crossing point is with an electromagnetic luminosity of $L_{\rm em}=10^{54}\,(\Gamma/220)^4$ erg\,s$^{-1}$. Since the observed peak gamma-ray luminosity is just $L_\gamma\sim10^{54}$ erg\,s$^{-1}$, in the region below the green dashed line, pairs can cool quickly by IC scattering prompt MeV photons. However, for $\Gamma\ga220$, $L_{\rm em}>L_\gamma$ and synchrotron cooling must contribute. For $\Gamma\ga400$ (above the solid green line), even $L_{\rm em}\ga10^{55}$erg\,s$^{-1}$ is needed for fast cooling.

Note, the Lorentz factor of GRB 221009A has been derived to be large, $\Gamma>400$. For example, the measured time delay of TeV relative to keV-MeV emission in GRB 221009A provides a direct probe to the Lorentz factor, $\sim 740\,(\eta_\gamma A_{0})^{-1/8}$ for interstellar medium or $\sim 680\,(\eta_\gamma A_{34})^{-1/4}$ for stellar wind medium, where the radiation efficiency $\eta_\gamma$ and the parameters on medium density $A_0$ and $A_{34}$ are of order unity \citep{2024arXiv240403229Z}.
Thus, the line emission region of GRB 221009A is likely to be with $L_{\rm em}\simeq L_B \ga10\, L_\gamma$.

If $f_{\rm v}=1$, the condition of fast pair annihilation (red dashed line; Equation \eqref{eq:ann}) seems to give large optical depth for Thomson scatterings off pairs, since the two processes have the same targets and comparable cross sections. However, the fast pair annihilation also reduces the time that a photon can interact with pairs to $0.1t_{\rm dyn}$. The optical depth of Thomson scatterings off pairs is, then,
\begin{equation}
    \tau_{\rm es}\simeq 0.1\langle n_\pm' \rangle \sigma_T r/\Gamma\simeq (8/30)f_{\rm v} \,t'_{\rm dyn}/t'_{\rm ann}\sim(8/3) f_{\rm v}. 
    \end{equation} 
Still, to obtain an optically thin emission region ($\tau_{\rm es}<1$), $f_{\rm v}\la3/8$ is required, corresponding to the parameter space above the red solid line in Figure \ref{fig:r-Gamma}, i.e., $\Gamma\ga133 \,(\delta E_\gamma/{\rm 10^{54}erg})^{1/(5+2\alpha)}(L_\gamma/{\rm 10^{54}erg\,s^{-1}})^{1/(5+2\alpha)}$. Thus the fast annihilation requires that the pair annihilating region is slightly clumpy with $f_{\rm v}\la3/8$, and the pair annihilation rate is enhanced somewhat by the slightly larger pair density.

In the internal shock model for kinetic-energy-dominated jets, the internal collisions occur at $r_{\rm IS}\simeq 2\Gamma^2ct_{v}\sim1.2\times10^{15}{\rm cm}~(\Gamma/500)^2(t_{v}/0.082~{\rm s})$, where $t_v\sim0.082$s is the rapid variability time observed in GRB 221009A  \citep{Liu2023}. This small radius is in contradiction with the constraint of the radius of the line emission region. However, in the magnetic-energy-dominated jet model, a large radius ($\ga10^{16}$ cm) for the energy dissipation region is expected \citep[e.g.,][]{2003astro.ph.12347L,2011ApJ...726...90Z}. Moreover, if $\Gamma>400$ in GRB 221009A, as derived by, e.g., \cite{2024arXiv240403229Z}, the fast cooling of pairs requires $L_{\rm em}\gg L_\gamma$. This further implies a magnetic luminosity $L_B\gg L_\gamma$, and thus, the jet is likely magnetic-energy dominated. This is consistent with the implication from the other reasoning for GRB 221009A \citep[e.g.,][]{2023SCPMA..6689511W}.

The observed decaying power-law index $k=1$ of the central energy of the emission line indicates a constant $r/\Gamma$ for the line emission region. Since $\Gamma\gg1$ always gives $\beta\approx1$, $r$ is constantly independent of angle ($\theta$) for a certain source frame time; hence, it is more likely that both $\Gamma$ and $r$ are independent of angle. Thus the deviation of the line flux decaying law from $k=3$ is likely due to angle dependence of line emission, rather than dynamics.

The last two data points of the emission line show somewhat of a deviation from the trend (Figure \ref{fig:evo}). If true, then we cannot exclude that some physics gets into play to account for this deviation.

The constraint of the line emission region in Equation \eqref{eq:variation1} may apply to the prompt gamma-ray emission region as well because the pair formation and cooling are undergoing at the same time. This means that the prompt gamma-ray emission may occur in a timescale (in either a comoving or source frame) 10 times shorter than the dynamic time. 

\subsection{The Second Bump of Prompt Emission}
In the analysis above, we have considered that the line emission is related to the first bump of the prompt emission, but one may ask why the second bump seems not contribute to the line emission. Conceivably, there may be several possible reasons. First, the second bump shows lower gamma-ray luminosity, which may not satisfy the conditions of efficient cooling in Equation \eqref{eq:rcSR1/IC1}; thus, no narrow, bright line emission could show up. Even if pairs annihilated, the relative large velocity leads to large broadening of the emission line, making it indistinguishable from the continuum spectrum. Second, the annihilation emission line may lie in higher energy, beyond the energy range of the detector. When it decreased into the MeV energy range, the line flux may be too low to be observed.

\subsection{Other Models}

One may expect to explain the power-law decaying behaviors of the emission line by the dynamical evolution of the GRB jet. However, such a scenario will face many difficulties. First, the bulk Lorentz factor of the jet-powered shock should follow $\Gamma\propto (t-t_0)^{-k}$, with $k=1$. For a medium-density $n$ distribution of $n\propto r^{-s}$, where $s\in[0,3)$, with $s=0$ and $2$ for homogeneous medium and wind medium, respectively \citep{DaiLu1998}, the shock dynamics is derived to be with $k=(3-s)/(8-2s)<3/8$.
Thus, it contradicts with the observed value of $k\approx1$. 
If $s\geq3$, the medium density will decrease steeply with radius and the shock even speeds up with $r$. To obtain $k\sim1$, one needs to introduce density bumps in the medium in front of the shock.
Moreover, in order to explain the line emission, the bulk Lorentz factor of the shock has decreased to $\Gamma\sim20$ within $\sim100$ s. The too low $\Gamma$ is in contradiction with the TeV emission modeling  \citep{2024arXiv240403229Z,LHAASO23}. Finally, in the external-shock region with radius beyond the deceleration radius $r_{\rm dec}\sim10^{17}~{\rm cm}~(E_{\rm k}/10^{55}~{\rm erg})^{1/3}(\Gamma_{0}/440)^{-2/3}(n/1~{\rm cm}^{3})^{-2/3}$, the born pairs are unable to reach an NR state because the luminosity of the external-shock emission is too low to satisfy Equation \eqref{eq:rcSR1/IC1}.

One may seek an alternative mechanism to the pair annihilation line. First of all, the spectral feature may not be accounted for by the thermal component, because the blackbody-like spectral profile will show a much broader profile, incompatible with the observed narrow line width.
Alternative line emission mechanisms could also be considered.
For instance, within our framework, if the MeV emission line is actually associated with the fluorescent iron line at $\sim6.7$ keV, the line emission is produced at $r\sim3\times10^{18}\,{\rm cm}~(\Gamma/500)$, an unreasonably large distance. 
Recently,  \cite{Wei:2024nbv} proposed a keV-scale atomic line of a heavy element boosted by a jet with $\Gamma\sim10^3$ to account for the multi-MeV lines reported by \cite{Maria2023}. However, the model needs to adjust for the more comprehensive observational results reported by \cite{YQ2024}, e.g., requiring $\Gamma\gg10^3$ for the observed line central energy up to about 37 MeV and the explanation for the power-law evolution of the line.

If the MeV line is directly from the nuclear decay of a specific radioisotope (mass number $A_{\rm iso}$) synthesized during this GRB, 
then there may be problems with finding a feasible nucleosynthesis process to create such a radioisotope with a short lifetime and excessively large mass corresponding to the measured data. Observation requires the mean lifetime of the radioisotope $\tau_d$ to cover the whole observation duration $t_{\rm dur}\sim135$ s, i.e., $\tau_d\gtrsim\Gamma\, t_{\rm dur}\gtrsim10^3-10^4$s provided $\Gamma\sim E_{\rm line}/E_d\sim10-100$, as the nuclear decay line energy $E_d\sim 0.1-3$ MeV. 
Moreover, the measured line luminosity $L_{\rm line}=\Gamma^2f_d E_d M_{\rm iso}/(\tau_d A_{\rm iso} m_b) \sim10^{51}$erg\,s$^{-1}$ requires $M_{\rm iso}\ga A_{\rm iso}{f_d}^{-1} M_\odot$, where $f_d\lesssim1$ is the branching ratio of the nuclear line, $m_b$ is the baryon mass, and $M_\odot$ is the Solar mass. Such an amount is orders of magnitude more abundant than the typical mass of the individual radioisotope synthesized in an explosive event associated with GRBs. For example, a traditional core collapse supernova is expected to dominantly produce $\lesssim 0.1 M_\odot$ $^{56}{\rm Ni}$ with $\tau_d\sim 10^6s$.

\section{Summary}
\label{sec:sum}

We provide a model to interpret the recent observational results of the MeV emission line in GRB 221009A, including the maximum line energy up to 37 MeV, the power-law time evolving of line central energy and flux, and the narrow line width, and hence probe the GRB jet physics with the line emission properties. In our model, electron-positron pairs are generated in prompt emission due to absorption of the highest energy photons; pairs should cool quickly down to the NR state and annihilate each other, producing the line emission. The high-latitude emission from the line emission region naturally gives rise to the temporal decrease of the emission line central energy with a power-law index $-1$.

In summary, several points are made from the observed line properties:

(1) The central energy evolution and the line luminosity constrain the radius of the emission region to be $r\ga10^{16}$cm. 

(2) The narrow line width $\sigma/E_{\rm line}\sim10\%$ requires that the line emission, as well as the prompt continuum emission, occurs within 10\% of the dynamical time. The implications include that pairs must cool fast down to the NR state within 10\% of the dynamical time. This means that, if $\Gamma\ga400$, then $L_{\rm em}\sim L_B\gg L_\gamma$ in GRB 221009A. Given $\Gamma>400$ derived from the TeV data modeling \citep{2024arXiv240403229Z}, the GRB 221009A jet is expected to be magnetic-energy dominated rather than kinetic-energy dominated.

(3) The requirement of small optical depth for Thomson scatterings off pairs $\tau_{\rm es}<1$ derives $\Gamma\ga130$. 

(4) The requirement of short pair annihilation time ($t'_{\rm ann}<0.1t'_{\rm dyn}$), together with that of small optical depth ($\tau_{\rm es}<1$), implies that the spatial distribution of annihilating pairs should be slightly clumpy with filling factor of $f_{\rm v}\la3/8$.

(5) The observed power-law decay index of line flux, $k\approx2$, deviated from the expected value of $k=3$ in simple high-latitude effects, suggests some angle dependence of line emission. 

(6) As for other scenarios, the external-shock origin and the emission mechanisms of MeV lines, rather than pair annihilation, all face many critical problems.

\section*{acknowledgments}
We appreciate the anonymous reviewer for helpful comments and suggestions. This work is supported by the International Partnership Program of Chinese Academy of Sciences (Grant No.113111KYSB20190020), the National Program on Key Research and Development Project (Grant No. 2021YFA0718500) from the Ministry of Science and Technology of China, the National Natural Science Foundation of China (Grant No. 12273042), the Strategic Priority Research Program of the Chinese Academy of Sciences (Grant Nos. XDA15360000, XDA30050000, and XDB0550300), and the funding from the Institute of High Energy Physics (Grant No. E25155U1) and the Chinese Academy of Sciences (Grant Nos. E329A3M1 and E3545KU2).

\bibliography{draft}
\bibliographystyle{aasjournal}





\setcounter{equation}{0}
\setcounter{figure}{0}
\setcounter{table}{0}
\setcounter{section}{0}
\setcounter{page}{1}
\makeatletter
\renewcommand{\theequation}{S\arabic{equation}}
\renewcommand{\thefigure}{S\arabic{figure}}
\renewcommand{\thetable}{S\arabic{table}}


\end{document}